\documentclass[fleqn,usenatbib]{mnras}

\usepackage{newtxtext,newtxmath}                 

\usepackage[T1]{fontenc}
\usepackage{ae,aecompl}

\usepackage{graphicx}	
\usepackage{amsmath}	
\usepackage{amssymb}	
\usepackage{epstopdf} 

\newcommand{\kms}{\,km\,s$^{-1}$}                          

\title[The dust making WR star HD 38030 in the LMC]
      {The episodic dust-making Wolf-Rayet star HD 38030 in the Large Magellanic Cloud.}

\author[P. M. Williams et al.]{
Peredur M. Williams,$^{1}$\thanks{E-mail: pmw@roe.ac.uk}
Nidia I. Morrell,$^{2}$ 
Konstantina Boutsia,$^{2}$
and Philip Massey$^{3,4}$
\\
$^{1}$Institute for Astronomy, University of Edinburgh, Royal Observatory, Edinburgh 
EH9 3HJ, UK\\
$^{2}$Las Campanas Observatory, Carnegie Observatories, Casilla 601, La Serena, Chile\\
$^{3}$Lowell Observatory, 1400 W Mars Hill Road, Flagstaff, AZ 86011, USA\\
$^{4}$Department of Astronomy and Planetary Science, Northern Arizona University, 
Flagstaff, AZ 86011-6010, USA \\
}
\date{Accepted 2021 June 3.
      Received 2021 June 3;
      in original form 2021 May 7.}

\pagerange{\pageref{firstpage}--\pageref{lastpage}}
\pubyear{2021}

\begin{document}

\maketitle

\label{firstpage}

\begin{abstract}
Mid-infrared photometry of the Wolf-Rayet star HD~38030 in the Large 
Magellanic Cloud from the NEOWISE-R mission show it to have undergone a 
dust-formation episode in 2018 and the dust to have cooled in 2019--20. 
New spectroscopy with 
the MagE spectrograph on the Magellan I Baade Telescope in 2019 and 2020 
show absorption lines attributable to a companion of type near O9.7III-IV. 
We found a significant shift in the radial velocity of the \ion{C}{iv} 
$\lambda\lambda$5801--12 blend compared with the RVs measured in 1984 and 1993. 
The results combine to suggest that HD~38030 is a colliding-wind binary 
having short-lived dust formation episodes, like the Galactic systems 
WR\,140 and WR\,19, but at intervals in excess of 20~yr.

\end{abstract}


\section{Introduction}
\label{Sintro}

It has long been known from infrared (IR) observations \citep*{ASH} that 
some Galactic WC-type Wolf-Rayet (WR) stars make carbon dust in their winds. 
Most of these WR dust makers do so persistently and are of spectral sub-types 
WC8--9 \citep*{WHT} but a few WR dust makers are of earlier subtypes and make 
their dust in brief, but regular, episodes at intervals of the order of a decade. 
The latter stars have been shown to be members of massive binary systems having 
highly elliptical orbits wherein the episodes of dust formation coincide with 
periastron passage, e.g. WR\,140 \citep{W90} and WR\,19 \citep{W19orb}. 
The dust formation is believed to result \citep{Usov91} from the collision of 
the carbon-rich WR stellar wind with that of its O-type companion, a process 
favoured when the stars are closer to each other in their orbit, with the 
result that the collision occurs in the denser regions of the stellar winds. 
 
In a search for more episodic or variable dust-making Wolf-Rayet systems, a 
sample of WC stars in the Galaxy and Large Magellanic Cloud (LMC) was 
monitored in the mid-IR using observations made in the first five years 
of the Near-Earth Object {\em WISE} Reactivation (NEOWISE-R) mission 
\citep{NEOWISER}, a reactivation of the {\em WISE} \citep{WISE} mission. 
Amongst the WC stars found to show variable dust emission \citep{WRNEOWISE} 
was HD~38030\footnote{Also WS~39 \citep{WSLMC},  Br~68 \citep{Brey} and 
BAT99~84 \citep*{BAT99}.} in the LMC: in 2018, after some years of apparent 
constancy in the IR, its $W1$ (3.4-\micron) and $W2$ (4.6-\micron) fluxes 
brightened by 1--2~magnitudes and its $W1-W2$ colour became more 
characteristic of emission by heated dust. 

This made HD~38030 the second WR system in the LMC to show dust formation, 
after the variable (persistent) dust-maker HD~36402 \citep{W36402}.
Recently, \citet{Lau21} identified six candidate WC-type dust-makers from 
their mid-IR flux variations in low metallicity extragalactic environments 
-- NGC~604 on the outskirts of M~33, NGC~2403, M~101, NGC~6946 and IC~1613 
-- showing this phenomenon to be widespread.

The fast rise in the IR flux from HD~38030 resembled those of WR\,140 and 
WR\,19, suggesting that HD~38030 might also be a collidng-wind binary (CWB) 
having a high orbital eccentricity. 
This prompted us to obtain spectra to search for evidence of binarity. 
The spectrum of HD 38030 had been classified as WC5+OB by \citet{SmithNB}, 
but the star has not been confirmed to be a binary. From the lack of 
variability in its radial velocities (RVs) observed in 1984 and 1993, 
\citet*{Bartzakos} deduced that it was `almost certainly' a single star.
Although the IR photometric history implies a long period and hence a small 
RV amplitude if the system is a CWB, the variation in RV might have been 
greater near IR maximum if that had occurred close to periastron passage, 
as in WR\,140 and WR\,19. This prompted our first spectroscopic observation 
in 2019 May, soon after the 2018 NEOWISE-R data showing the dust formation 
became available, and further observations in 2020.
Also, as the NEOWISE-R mission continued, we sought to monitor the dust 
emission with further mid-IR observations after its rise with a view to 
determining the character and duration of the episode and estimating the 
time-scale and possible periodicity of the outbursts.

\section{Observations}
\subsection{Spectroscopy}
Spectra of HD 38030 were observed with the MagE (Magellan Echellette) 
Spectrograph \citep{MagE} on the Magellan I Baade telescope at Las Campanas 
Observatory on 2019 May 19 and 2020 January 15 and November 26.
We used the 1-arcsec slit providing a resolving power of R $\sim$ 4100. 
A short exposure of a ThAr lamp obtained immediately after the target 
observation was used for wavelength calibration. Exposure times were 300~s 
for the 2019 spectrum and 200~s for the other two, producing S/N ratios of 
200-300 and 100-150, respectively.

Flux calibration was performed from spectrophotmetric standard spectra taken 
during the same observing nights. The data were processed with a combination 
of the {\sc iraf} echelle tasks and the {\it mtools} package originally 
designed by Jack Baldwin for the reduction of MIKE data (available from the 
LCO web-site).
The relative flux calibration of MagE is good to a few percent, as both 
spectrophotometric standards and program objects are observed at the 
parallactic angle.  The absolute fluxes are uncertain at the 30 per cent 
level or greater, as they depend upon similar slit losses for the program 
object and spectrophotometric standards.

\subsection{Photometry}

The IR photometric history of HD~38030 was assembled from a number of sources, 
the most useful being the {\em WISE}, NEOWISE Post-Cryo \citep{NEOWISE} and 
NEOWISE-R surveys from which the initial brightening in 2018 was discovered. 
The wavelengths of the {\em WISE} $W1$ and $W2$ bands, 3.4 and 4.6$\mu$m, 
are well placed for observing emission by $T_g \sim$ 1000-K dust in the 
wind of a hot star. Individual photometric class `AA' $W1$ and $W2$ magnitudes
were retrieved from the WISE All-Sky, WISE Post-Cryo and NEOWISE-R Single  Exposure (L1b) Source Tables provided by the NASA/IPAC Infrared Science 
Archive\footnote{{\tt https://irsa.ipac.caltech.edu}}. 
Observations were selected to be within 1 arc sec of the optical position;  
in practice, the median distance of the 1336 individual observations was found 
to be 0.24 arc sec.
The {\em WISE} data were collected in `visits', akin to observing runs, each 
including a number of observations taken at intervals of one or more 94-min 
orbital passages and spread over several days, with the visits separated by 
about six months as the Sun-synchronous orbit of the satellite followed the 
Earth in its orbit. 
The length of visit and number of observations in each depend on the overlap 
of the survey strips, which increase with increasing ecliptic latitude as 
the overlap of the survey strips on the sky increases. 
This privileges HD~38030 on account of its high ecliptic latitude (-86\degr): 
it was observed an average of 87 times in each visit, thereby increasing 
the accuracy of the mean magnitudes from each visit, which are the data we 
use to monitor long-term variability (Table~\ref{TWISE}). Fortuitously for 
the study of the dust formation episode, the first of the visits in 2018 and 
2019 were protracted and split by date, so that they could be treated as two 
separate runs, with the result that we have three well separated observations 
in each of 2018 and 2019. The data from the {\em WISE} surveys from the 2021 
data release are collected in Table~\ref{TWISE}.

Earlier mid-IR observations of HD 38030 were taken during 2005 in the SAGE 
\citep{SAGEphot} survey. These data are given in Table~\ref{TSAGE}. 
In between the second SAGE and first of the {\em WISE} observations, HD~38030 
was also observed at 3.2, 7.0 and 11.0$\micron$ in the AKARI Infrared Camera 
Survey of the Large Magellanic Cloud \citep{AKARI} on 2006 October 27--30. 
These data are given in Table~\ref{TAKARI}.

In the near-IR, HD~38030 was observed in 1996 and 1997 in the Deep Near 
Infrared Survey of the Southern Sky \citep[DENIS,][]{DENIS} Survey, 3rd 
Data Release \citep{DENIS3}; in 2000 in the Two-micron All-Sky Survey 
\citep[2MASS,][]{2MASS}; in 2000 and 2001 in the deeper 2MASS 6X w/LMC/SMC 
Point Source Working Database \citep{LMC6X}; in 2003 in the InfraRed 
Survey Facility (IRSF) Magellanic Cloud Survey \citep{IRSF} and very 
frequently (46 times) during 2009--10 in the Vista Magellanic Clouds Survey 
\citep[VMC,][]{VMC1}. The $K_s$ magnitudes from the DENIS, 2MASS and IRSF 
surveys are collected in Table~\ref{TKs}. For brevity, we do not list all 
the VMC data but give only the average $K_s$ from the first and last nights 
of the 2009--10 intensive sequence of observations, followed by that from 
the final night later in 2010, all retrieved from the VMCDR4 detection table 
in the VISTA Science Archive\footnote{{\tt http://horus.roe.ac.uk/vsa}} \citep{VSA}.

\begin{table} 
\centering 
\caption{SAGE IRAC Epoch 1 and Epoch 2 magnitudes of HD~38030; 
the dates were recovered from the {\em Spitzer} Heritage Archive.} 
\label{TSAGE}
\begin{tabular}{ccccc}
\hline
  Date  & [3.6]          & [4.5]          & [5.8]          &  [8.0]         \\
\hline
2005 55 & 12.80$\pm$0.04 & 12.53$\pm$0.05 & 12.44$\pm$0.04 & 12.19$\pm$0.08 \\
2005.83 & 12.81$\pm$0.06 & 12.60$\pm$0.07 & 12.42$\pm$0.06 & 12.17$\pm$0.07 \\
\hline
\end{tabular}
\end{table}

\begin{table}
\centering
\caption{Magnitudes of HD~38030 in the AKARI LMC Survey.}
\label{TAKARI}
\begin{tabular}{lccc}
\hline
Date    &   $N3$         &   $S7$         &  $S11$        \\
\hline
2006.83 & 12.71$\pm$0.03 & 12.27$\pm$0.04 & 11.93$\pm$0.08 \\
\hline
\end{tabular}
\end{table}

\begin{table} 
\centering 
\caption{{\em WISE} photometric history of HD~38030 from the several surveys. 
The dates are the average dates of each visit.} 
\label{TWISE}
\begin{tabular}{cccl}
\hline
Date  &    $W1$        &    $ W2$       & Survey  \\
\hline        
2010.31 & 12.68$\pm$0.02 & 12.48$\pm$0.02 & All-Sky   \\
2010.80 & 12.72$\pm$0.02 & 12.48$\pm$0.02 & Post-Cryo \\
2014.32 & 12.69$\pm$0.01 & 12.53$\pm$0.01 & NEOWISE-R \\
2014.82 & 12.70$\pm$0.01 & 12.51$\pm$0.01 & NEOWISE-R \\  
2015.31 & 12.71$\pm$0.01 & 12.52$\pm$0.01 & NEOWISE-R \\  
2015.80 & 12.70$\pm$0.01 & 12.52$\pm$0.01 & NEOWISE-R \\  
2016.31 & 12.71$\pm$0.01 & 12.54$\pm$0.01 & NEOWISE-R \\  
2016.80 & 12.70$\pm$0.01 & 12.54$\pm$0.01 & NEOWISE-R \\  
2017.31 & 12.70$\pm$0.01 & 12.53$\pm$0.01 & NEOWISE-R \\  
2017.77 & 12.71$\pm$0.01 & 12.54$\pm$0.01 & NEOWISE-R \\   
2018.31 & 12.00$\pm$0.01 & 11.26$\pm$0.01 & NEOWISE-R \\   
2018.43 & 11.87$\pm$0.02 & 11.05$\pm$0.02 & NEOWISE-R \\   
2018.76 & 11.52$\pm$0.01 & 10.62$\pm$0.01 & NEOWISE-R \\   
2019.31 & 11.77$\pm$0.01 & 10.85$\pm$0.01 & NEOWISE-R \\   
2019.40 & 11.89$\pm$0.02 & 10.97$\pm$0.03 & NEOWISE-R \\   
2019.77 & 12.17$\pm$0.01 & 11.33$\pm$0.01 & NEOWISE-R \\
2020.31 & 12.45$\pm$0.01 & 11.81$\pm$0.01 & NEOWISE-R \\
2020.76 & 12.60$\pm$0.01 & 12.08$\pm$0.01 & NEOWISE-R \\   
\hline
\end{tabular}
\end{table}

\begin{table} 
\centering 
\caption{Near-IR $K_s$ photometric history of HD 38030. The DENIS 
data are psf magnitudes from the third release \citep{DENIS3}. 
The VMC entries are averages of the three observations in each 
of the first and last nights of the intensive series and the 
final night later in 2010.} 
\label{TKs}
\begin{tabular}{lcl}
\hline
Date    & $K_s$          &  Survey \\
\hline
1996.97 & 12.56$\pm$0.13 & DENIS DR3 \\ 
1997.15 & 12.70$\pm$0.16 & DENIS DR3 \\ 
2000.10 & 12.96$\pm$0.04 & 2MASS     \\
2000.94 & 12.86$\pm$0.03 & 2MASS6X   \\ 
2001.10 & 12.88$\pm$0.03 & 2MASS6X   \\ 
2003.86 & 12.90$\pm$0.02 & IRSF      \\ 
2009.84 & 12.90$\pm$0.01 & VMC DR4   \\ 
2010.19 & 12.90$\pm$0.01 & VMC DR4   \\
2010.86 & 12.89$\pm$0.01 & VMC DR4   \\
\hline
\end{tabular}
\end{table}

\section{Results}
\subsection{The spectrum}
%
\begin{figure*}                                                
\centering
\includegraphics[angle=270,width=17cm]{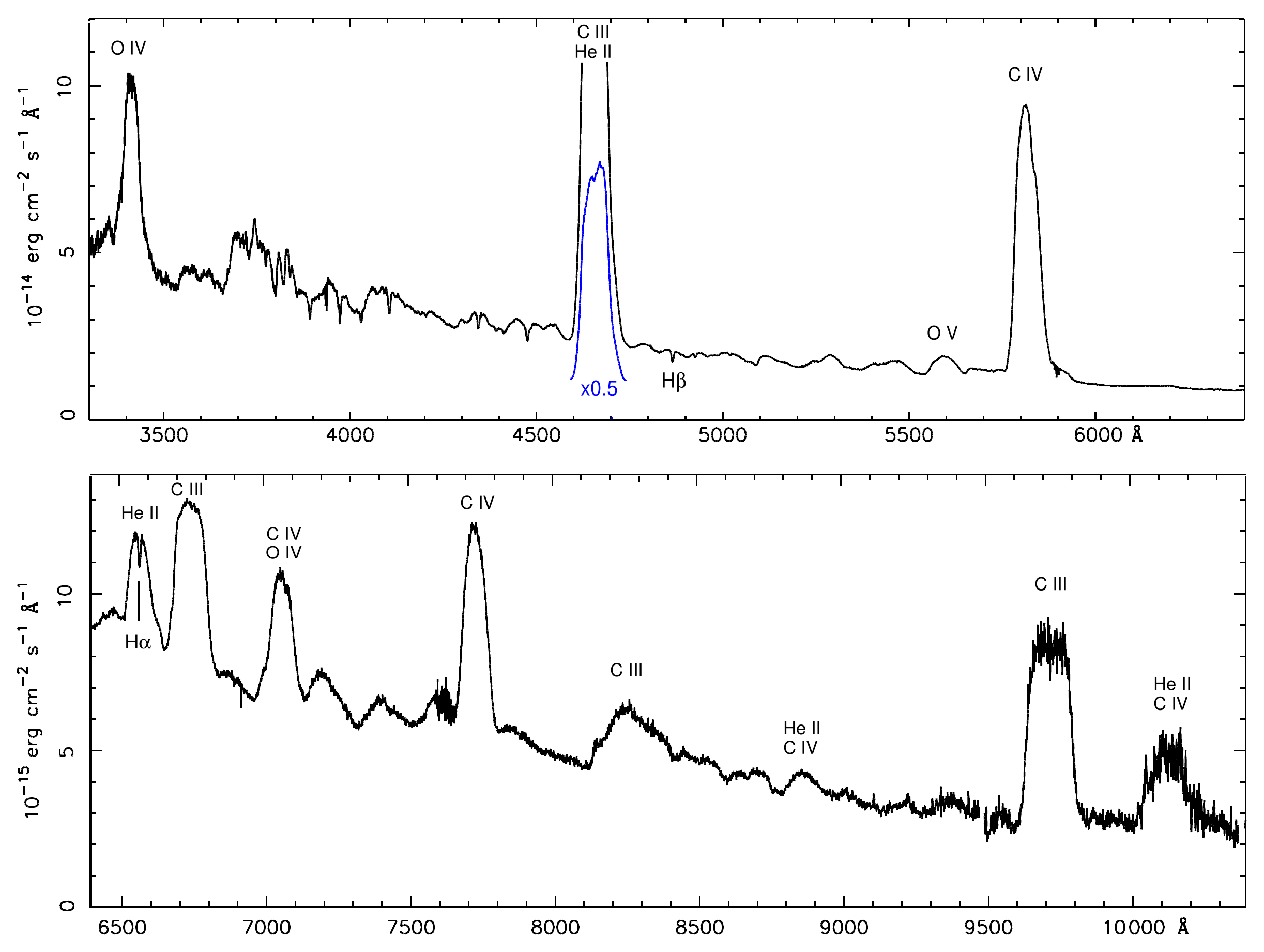}                        
\caption{Optical spectrum of HD~38030. Note the 10-fold difference in flux 
density between the panels. To provide a useful scale, the strong 4650-\AA\ 
\ion{C}{iii}+\ion{He}{ii} feature was truncated and a half-scale copy inserted 
in colour.}
\label{Fsp}
\end{figure*}                                 

\begin{figure}                                                
\centering
\includegraphics[angle=270,width=8.5cm]{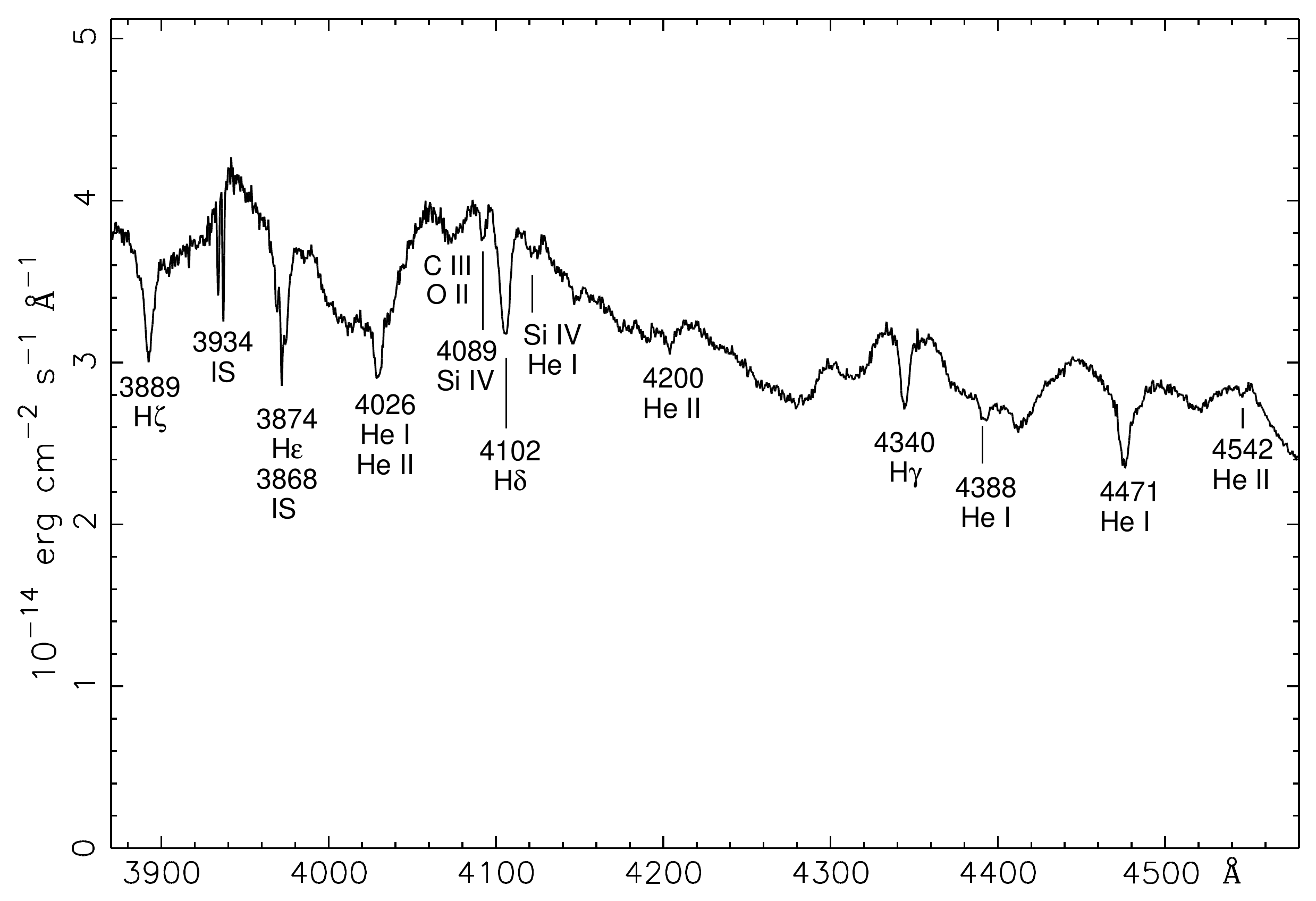}                        
\caption{Higher scale spectrum of HD~38030 in the blue showing absorption lines.}
\label{FAbs}
\end{figure}                                 

The 3300--10400-\AA\ spectrum of HD 38030 observed on 2019 May 19 is shown 
in Fig.\,\ref{Fsp}. The emission-line spectrum resembles that of the LMC WC4 
star HD~37026 \citep{CrowtherWC4} with strong \ion{He}{ii}, \ion{C}{iii-iv} 
and \ion{O}{iv-v} emission lines but with the significant difference that 
HD~38030 shows absorption lines. 
Superimposed on the $\lambda$6560 \ion{He}{ii} emission we see H$\alpha$ in 
absorption. Also visible in Fig.\,\ref{Fsp} is H$\beta$ absorption.
The higher Balmer, \ion{He}{i-ii} and \ion{Si}{iv} absorption lines are 
more easily seen in the higher scale spectrum in Fig.\,\ref{FAbs}. 
The absorption lines point to the presence of an OB type companion. 

\begin{table} 
\centering 
\caption{Absorption lines in the spectrum of HD 38030 measured from spectra 
observed on 2019 May 19, 2020 January 15 and November 26, giving identifications 
(including contributing blends), observed wavelengths and equivalent widths (EW).  Wavelengths measured by bisection and less certain data are marked with colons (:).}
\label{TAbs}
\begin{tabular}{lllclclc}
\hline
\multicolumn{2}{c}{Identification} & \multicolumn{2}{c}{2019 May} & \multicolumn{2}{c}{2020 Jan.} & \multicolumn{2}{c}{2020 Nov.}\\
  Ion        & lab.wl. &   wl.  &  EW   &    wl. &  EW  &  wl.  & EW \\   
             & (\AA)   & (\AA)  & (\AA) &  (\AA) & (\AA)& (\AA) & (\AA)\\     
\hline
H$_{11}$     & 3770.63 & 3774.1 & 0.4  & 3774.4 & 0.6  & 3774.2 & 0.5 \\
H$\eta$      & 3835.40 & 3838.4 & 0.5  & 3838.4 & 0.4  & 3838.5 & 0.4 \\
H$\zeta$     & 3889.06 & 3892.9 & 0.8  & 3892.8 & 0.9  & 3892.7 & 0.9 \\
+ \ion{He}{i}&         &        &      &        &      &        &     \\
H$\epsilon$  & 3970.08 & 3974.2:& ...  & 3974.3:& ...  & 3974.2 & ... \\
bl IS        &         &        &      &        &      &        &     \\
H$\delta$    & 4101.74 & 4105.1 & 1.4  & 4105.6 & 1.4  & 4104.9 & 1.4\\
+ \ion{N}{iii} &       &        &      &        &      &        & \\
H$\gamma$    & 4340.46 & 4344.5 & 0.8  & 4343.9 & 1.0  & 4344.6 & 0.8 \\
H$\beta$     & 4861.32 & 4865.5 & 1.0  & 4865.6 & 1.0  & 4865.4 & 1.1 \\
H$\alpha$    & 6562.80 & 6568.1 & 0.7  & 6568.4 & 0.7  & 6568.2 & 0.6 \\
\ion{He}{i}  & 4471.47 & 4475.9 & 1.0  & 4476.2 & 0.9  & 4476.0 & 0.9 \\
\ion{He}{i}  & 4026.19 & 4029.6 & 0.5  & 4029.7 & 0.5  & 4029.7 & 0.5 \\
+ \ion{He}{ii} &       &        &      &        &      &        & \\
\ion{He}{i}  & 4387.93 & 4392.0 & 0.2: & 4391.7:& ...  &  ...   & ... \\
\ion{He}{i}  & 4921.93 & 4926.1 & 0.4  & 4926.8 & 0.3  & 4926.7 & 0.4 \\
\ion{He}{ii} & 4199.83 & 4203.6 & 0.1  &  ...   & ...  &  ...   & ... \\
\ion{He}{ii} & 4541.59 & 4546.1 & 0.1: &  ...   & ...  & 4545.9 & 0.1:\\
\ion{He}{ii} & 5411.52 & 5416.3 & 0.2  &  ...   & ...  & 5416.3 & 0.1:\\
\ion{Si}{iv} & 4088.86 & 4092.5 & 0.2  &  ...   & ...  & 4092.7 & 0.1:\\
\hline
\end{tabular}
\end{table}

The wavelengths and equivalent widths (EWs) of the absorption lines are 
given in Table~\ref{TAbs}. The EWs were measured by direct integration 
over the profile using bespoke software, with a local continuum determined 
from fitting the flux at six points (spanning 1.2\AA) on either side of the 
line. The same code gave absorption-weighted wavelengths, 
$\int \lambda A_{\lambda} d\lambda / \int A_\lambda d\lambda $, for these 
lines where $A_{\lambda}$ is the difference between the continuum and 
residual flux at that wavelength.
Independent measurement of the EWs using {\sc iraf} routines gave 
values consistent within a fraction of 0.1\AA.
Where it was not possible to measure an EW because of blending or formation 
on the slope of an emission line, the wavelength was estimated by bisection 
of the profile. Some of the weaker lines could not be confidently measured 
in the 2020 spectra on account of their slightly lower S/N ratios.

The presence, but weakness, of the $\lambda$4542 \ion{He}{ii} 
line relative to $\lambda$4471 \ion{He}{i} indicates a 
spectral type near B0 for the absorption-line spectrum, no earlier 
than O9.5 or later than B0.2 \citep[cf.][]{OBatlas}. 
Detailed criteria for classification amongst the O8.5--B0 stars is 
provided by \citet{Sota2011}. 
While comparison with their \ion{He}{ii}\,/\,\ion{He}{i} ratios indicate 
O9.7 or B0, the weakness of $\lambda$4552 \ion{Si}{iii} relative to 
$\lambda$4542 \ion{He}{ii} in HD~38030 rules out B0 and favours O9.7 
or O9.5, leaving us to adopt O9.7 for the spectral type of the companion. 
As to luminosity, the usual classification lines for O type stars are 
masked by the very strong $\lambda$4650 WC4 emission feature and we have 
to consider the $\lambda$4089 \ion{Si}{iv}\,/\,$\lambda$4026 \ion{He}{i} ratio. 
Comparison of its value in HD~38030 ($\simeq$ 0.4) with those in O9.7 stars 
\citep[fig. 6]{Martins}, appears to rule out a supergiant classification and 
is most consistent with luminosity classes III or IV.

To estimate the dilution of the companion spectrum by the continuum of 
the WC4 star, we compared the EWs of four Balmer lines (H$\zeta$, 
H$\delta$, H$\gamma$ and H$\beta$) with those in late sub-type O stars. 
For the latter, we formed average EWs from the spectra of eight O9--O9.5 
stars in Lucke-Hodge 41 observed with the same instrumentation. 
The average of the ratios of the EWs (HD~38030/template) was 0.37$\pm$0.07, indicating that the companion was 1.1$\pm$0.2 mag fainter than the system 
(and 0.6$\pm$0.3 mag fainter than the WC4 star). 
For the system, we adopted $V = 12.99$ \citep*{Neugent_Census} and allowed 
for $E(B-V)$ = 0.19 of interstellar extinction. The latter was determined 
by following \citet{CrowtherWC4} in adopting a Galactic component of 
$E(B-V)$ = 0.07 from \citet{Schlegel} and adding a local component of 
$E(B-V)$ = 0.12 estimated from the relative strengths of the Galactic and LMC 
interstellar \ion{Ca}{ii} K lines in our spectra of HD~38030. For the absolute  magnitude, we use the distance to the LMC \citep[49.6 kpc,][]{LMCdist}, 
giving $M_V = -6.1$ for the system and $M_V = -5.0\pm0.2$ for the companion.
Comparison with luminosities of late sub-type O stars in the LMC 
\citep[e.g.][fig. 24]{Nolan2014} suggests that the class of the companion 
is in the region III-IV, possibly main-sequence but not supergiant.

\begin{figure}                               
\centering
\includegraphics[angle=270,width=8.0cm]{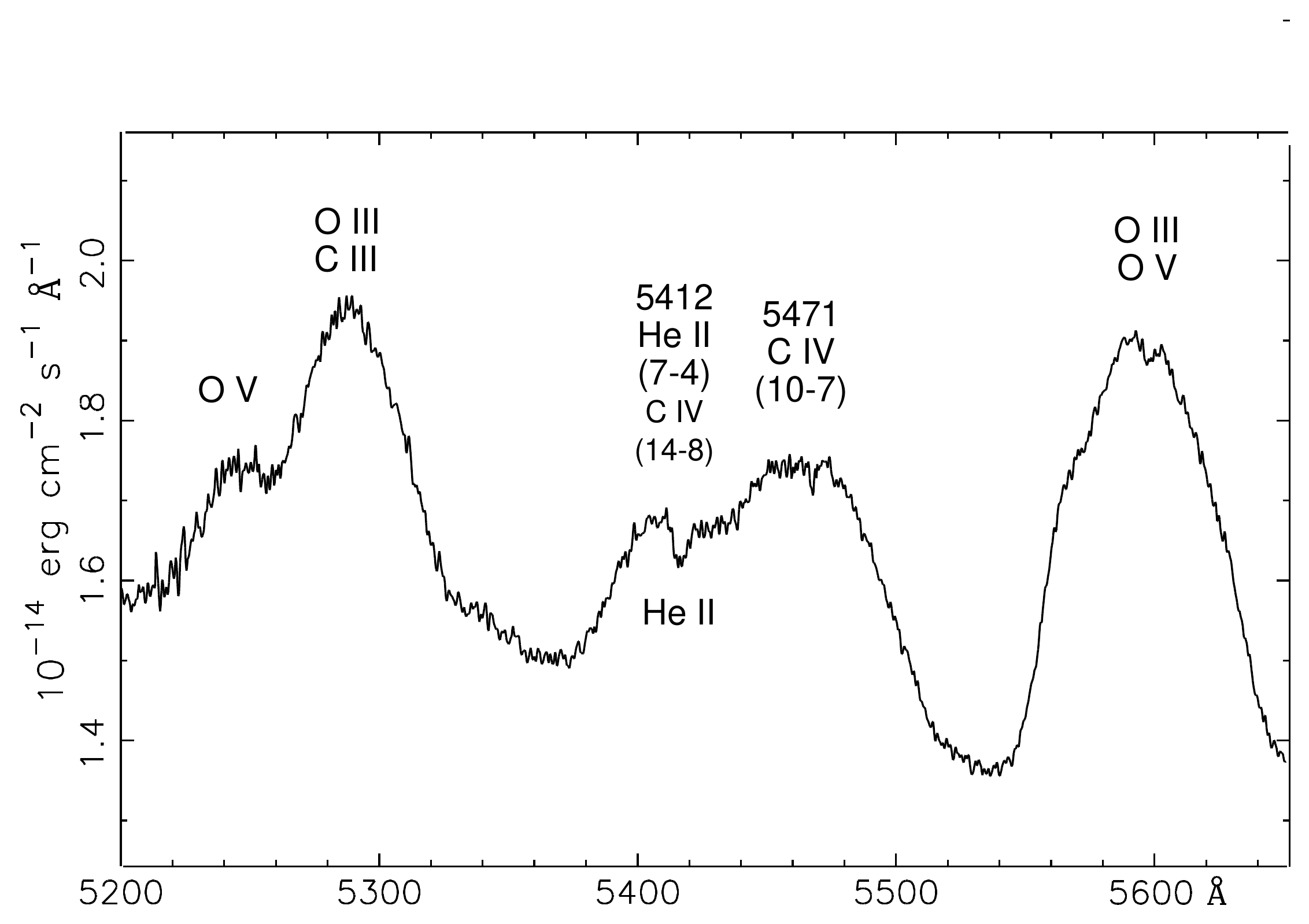}                        
\caption{Spectrum of HD~38030 in the region of the C/He abundance-sensitive 
$\lambda$5471 \ion{C}{iv} / $\lambda$5412 \ion{He}{ii} ratio. 
The \ion{He}{ii} line also shows an absorption line from the companion.}
\label{FCHe}
\end{figure}                                 

A model analysis of the emission-line spectrum is beyond the scope of this 
paper but we examined the C/He diagnostic used by \citet{CrowtherWC4}, 
the ratio of the $\lambda$5471 \ion{C}{iv} and $\lambda$5412 \ion{He}{ii}
intensities. This is shown in Fig.\,\ref{FCHe}. 
Comparison with the spectra of a sample of LMC WC4 stars in this region 
\citep[figs 2, 9]{CrowtherWC4}, 
bearing in mind the dilution of the HD~38030 emission-line spectrum by the 
companion, shows that the C/He abundance ratio in HD~38030 is at the high 
end of those in the sample of stars analysed by Crowther et al., having 
C/He $\simeq$ 0.35, which we suggest applies to HD~38030 as well.

\subsection{Search for radial velocity shifts.}
\label{SRVshift}

\begin{table}
\centering
\caption{Comparison of heliocentric radial velocities of the $\lambda\lambda$5801--12 \ion{C}{iv} blend determined by 
\citet[as Br~68]{Bartzakos} and from our spectra.}
\label{T5808}
\begin{tabular}{lll}
\hline
Date            &  RV (\kms)         &   Source  \\
\hline
1984 December   &  549$\pm$8.3  & 10 obs, Bartzakos et al. \\
1993 November   &  557$\pm$9.4  & 9 obs, Bartzakos et al. \\
2019 May        &  477$\pm$8    & this work \\
2020 January    &  486$\pm$8    & this work \\
2020 November   &  483$\pm$8    & this work \\
\hline
\end{tabular}
\end{table}

To test whether there had been any RV shift between the 1984--93 observations 
by \cite{Bartzakos} and our own, we aimed to replicate their RV measurements 
by measuring the intensity-weighted wavelength \citep[p. 26]{BartzakosThesis} 
of the strong $\lambda\lambda$5801--12 \ion{C}{iv} blend and deriving RVs 
adopting a laboratory wavelength of 5808.0~\AA. The errors on our RVs were 
found from repeated measurement of the line and are close to 0.1 of a MagE resolution element.
The heliocentric velocities are compared in Table~\ref{T5808}. 
It is evident that there is a real velocity shift, 77$\pm$12~\kms, between 
the two datasets. Assuming that this shift is not caused by variations 
in the shape of the \ion{C}{iv} emission-line profile, this supports the 
notion that the absorption-line spectrum forms in a physical companion to 
the WC4 star and the HD~38030 is also a colliding wind binary (CWB).

The difference between our 2019 and 2020 observations is comparable to 
the offset between the 2019 and 2020 interstellar line velocities 
(Table~\ref{TIS}) and is not considered significant. 
To look for a RV shift between our 2019 and 2020 spectra, we considered the 
absorption lines (Table~\ref{TAbs}). The average shift (2019 to 2020 November) 
from the stronger lines: H$\alpha$ to H$\zeta$, \ion{He}{i}  
$\lambda\lambda$ 4471 and 4025 is -2$\pm$9 \kms, so we did not detect any RV change in this interval -- which might have occurred if the stars were 
close to periastron in a highly elliptical orbit.

\begin{table*}
\centering
\caption{Wavelengths and heliocentric RVs of Galactic and LMC interstellar 
lines. Both Galactic and LMC IS components are measurable for the \ion{Ca}{ii} 
$H$ and $K$ lines but only one of each for the Na $D$ lines because the 
Galactic $D_1$ falls on to the LMC $D_2$.}
\label{TIS}
\begin{tabular}{lccccccc}
\hline
\multicolumn{2}{l}{Identification} & \multicolumn{2}{c}{2019 May} & 
\multicolumn{2}{c}{2020 Jan} & \multicolumn{2}{c}{2020 Nov.} \\
             &  (\AA)  & (\AA)  &(\kms)& (\AA)  &(\kms)& (\AA)  &(\kms)\\
\hline
\ion{Ca}{ii} & 3933.68 & 3934.0 &  22  & 3934.1 &  32  & 3934.1 & 30  \\
             &         & 3937.0 & 255  & 3937.1 & 261  & 3937.2 & 266 \\
\ion{Ca}{ii} & 3968.49 & 3969.0:&      & 3969.0:&      & 3969.1:&  \\
             &         & 3972.0:&      & 3972.1:&      & 3972.1:&  \\
\ion{Na}{i}  & 5889.95 & 5890.3 &  21  & 5890.3 &  19  & 5890.4 & 21 \\
\ion{Na}{i}  & 5895.92 & 5901.4 & 281  & 5901.5 & 284  & 5901.6 & 287 \\
\hline
\end{tabular}
\end{table*}

\subsection{The dust-formation episode}

\begin{figure}                                  
\centering
\includegraphics[angle=270,width=8.5cm]{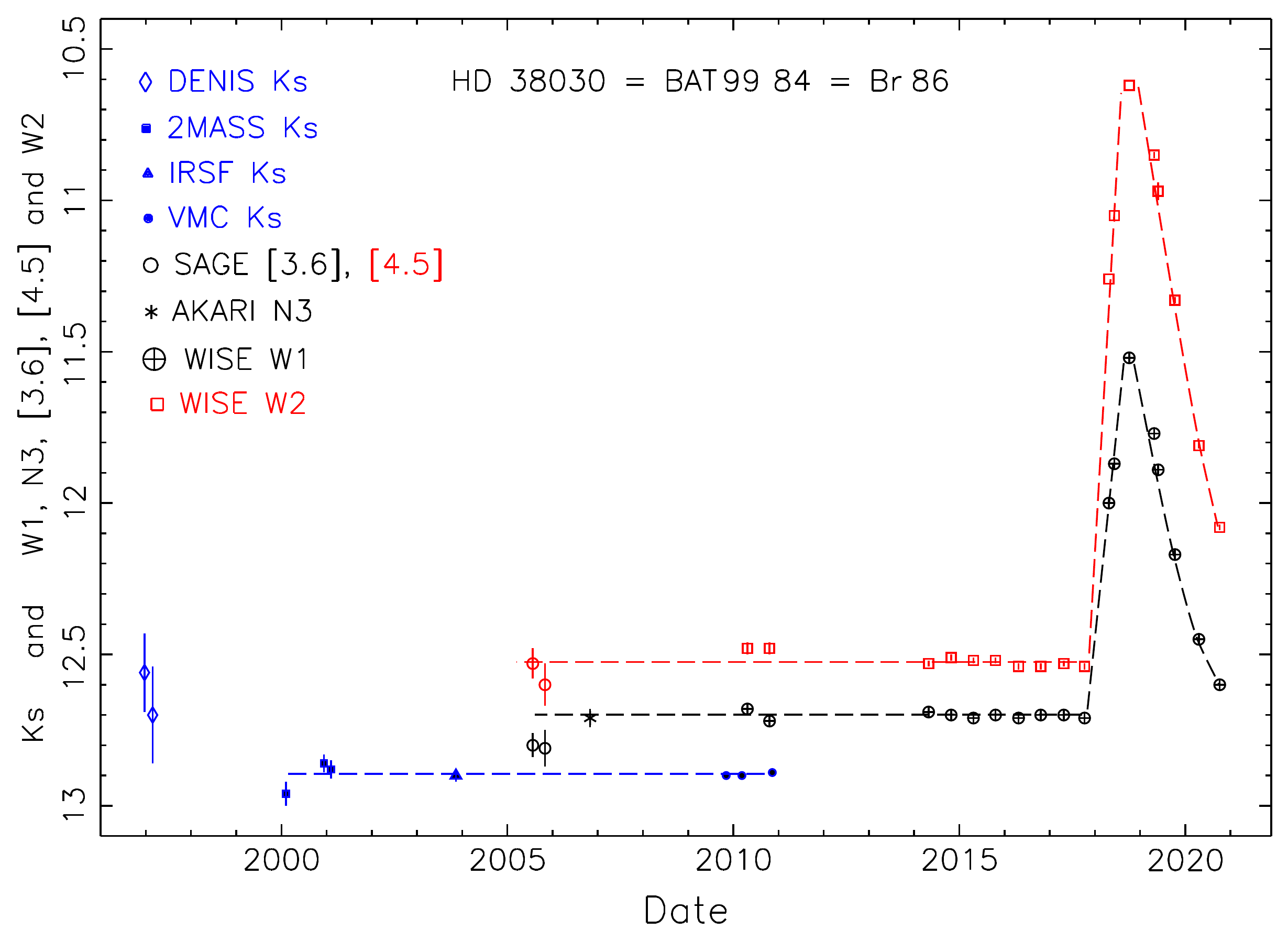}                        
\caption{Synoptic 3--5-\micron\ mid-IR and $K_s$ near-IR photometry of HD~38030. 
The dashed lines drawn through the data points do not represent any model.}
\label{Flc}
\end{figure}                                 

During 2010--2017, the {\em WISE} photometry of HD 38030 (Table~\ref{TWISE}, 
Fig.\,\ref{Flc}) showed no variation but, in 2018, the flux in $W1$ and $W2$ 
was observed to be rising sharply, at rates exceeding 1~mag~y$^{-1}$. 
We have no information about the onset of this event but, if the rate of 
brightening before our first 2018 observation was similar to that during 
the rise, formation must have begun soon after our last 2017 observation.
Nor do we know at what date after our last 2018 observations the emission 
reached its maximum but, from the fact that the individual values of $W1$ and 
$W2$ during the 18~days of that `visit' showed no significant brightening, 
the maximum must have occurred during or very soon after that observation. 
Examination of the positions of the 326 individual NEOWISE-R 
observations during the outburst shows no difference from the pre-outburst 
observations and a median distance of only 0.15 arc sec from the optical 
position, strongly identifying the outburst with the Wolf-Rayet system. 
An alternative eruptive source separated by $\lesssim$ 0.2 arc sec would also 
have fallen in the 1 arc sec slit used for the MagE spectrum but, as noted 
above, the latter (apart from the absorption lines) is very similar to the WC4 star HD~37026. There is no suggestion of extraneous features such as emission 
in the Balmer lines. If our sight line was in the Galactic plane, one might 
posit an IR eruption from a source in or behind a dense molecular cloud which 
extinguished the optical spectrum -- but the Catalogue of Molecular Clouds 
in the LMC \citep{LMC_GMC} has no molecular cloud within  2 arc min of 
HD~38030. We therefore reject this possibility and, recalling the positional 
coincidence, are confident in associating the IR outburst with the WC4 system.  

\begin{figure}                               
\centering
\includegraphics[angle=270,width=8.0cm]{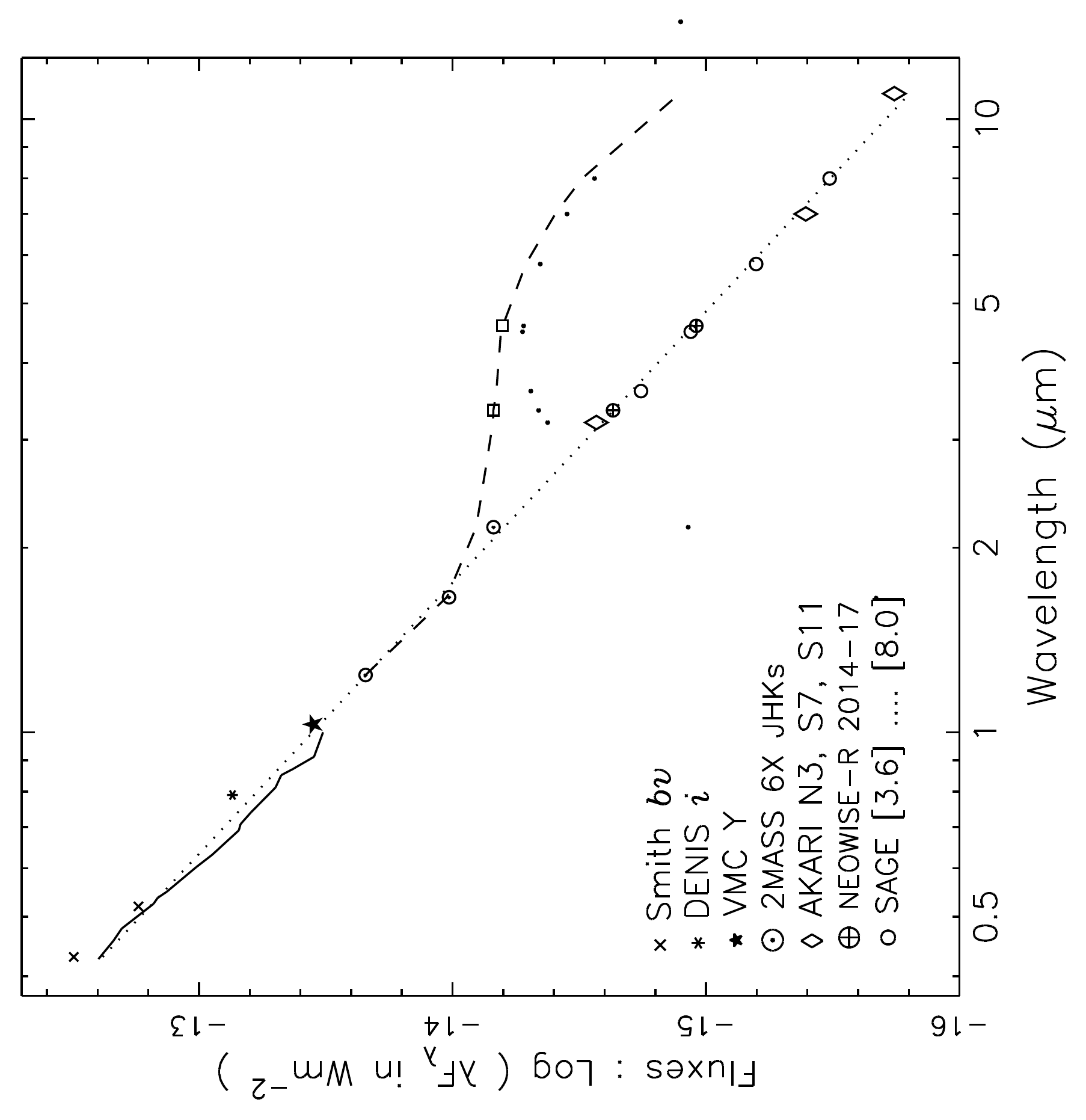}                        
\caption{Spectral energy distribution of HD~38030 showing emission by dust at 
IR maximum (fluxes ($\Box$) corresponding to $W1$ and $W2$ observed in 2018.76) 
together with the model isothermal dust cloud (small dots) and the dust plus 
wind (dashed line). The dotted line shows the wind SED defined by the 
multi-wavelength photometry, identified by different symbols.
The solid line in the optical region was formed from the line-free regions of 
our 2019 MagE spectrum.}
\label{FSED}
\end{figure}

In Fig.\,\ref{FSED}, the fluxes calculated from the 2018.76 values of $W1$ and 
$W2$ are plotted above the IR spectral energy distribution (SED) determined 
from the pre-outburst {\em WISE}, NEOWISE and NEOWISE-R data, together with 
the earlier {\em AKARI}, SAGE, 2MASS $JHK_s$ and VMC $Y$ data, which can be 
seen to be consistent. 
All data were de-reddened by $E(B-V)$ = 0.19 as above. 
The pre-outburst SED, taken to represent the stellar wind of HD~38030, can 
be represented by a power law, $\lambda F_{\lambda} \propto \lambda^{-2.26}$. 
Although not required for the present investigation, we note that extrapolation  
of this IR power law to optical wavelengths lies between the continuum 
from our MagE spectrum and fluxes from narrow-band $b$ and $v$ 
\citep{SmithNB}, DENIS $i$ and VMC $Y$ photometry also plotted in 
Fig.\,\ref{FSED}. The continuum estimated from the MagE spectrum, drawn 
avoiding emission lines, lies lower than the fluxes derived from the 
photometry owing to the inclusion of WR emission lines in the latter.

With fluxes at only two wavelengths during the episode of dust formation, we 
can not expect to see the Planckian IR SED characteristic of dust emission 
such as those shown by classsic WR dust makers \citep[e.g.,][]{CBK,WHT}, but 
the $W1$--$W2$ slope is much `redder' than that of the wind emission 
or of optically thin free-free emission\footnote{Such emission would have 
$W1$--$W2$ = 0.67, still `bluer' than the colours in 2018--19, and could affect 
the optical spectrum.}, so we consider the excess emission to be provided by 
heated dust. 

An increase of dust emission in the IR can be caused either by an 
increase in the stellar UV-visual flux heating the dust and being 
reprocessed in the IR or by the formation of new dust. The fact that the 
SED from the MagE spectrum lies no higher than that from the earlier 
photometry points to the latter alternative.
To get a measure of the amount of dust formed in the event, we use a simple 
model for isothermal, optically thin dust emission in which the flux, 
$F(\lambda)$ observed at a distance $d$ from the source is given by the 
convolution of the grain emissivity, $\kappa_{\lambda}$, and the Planck 
function for the grain temperature:

\begin{equation}
4\pi d^2 F_{\lambda} = m_{\rmn{g}} \kappa_{\lambda} B(\lambda, T_{\rmn{g}})
\end{equation}

Given $\kappa_{\lambda}$, the temperature of the grains, $T_{\rmn{g}}$, can 
be found by fitting the dust emission in $W1$ and $W2$. The mass of dust is 
then found using this temperature, the overall fit and distance, $d$, for 
which we adopted that to the LMC \citep[49.6 kpc,][]{LMCdist}.
We calculated $\kappa_{\lambda}$ from the optical properties of the `ACAR' 
amorphous carbon grains prepared in an inert atmosphere in the laboratory by 
\citet{ColangeliAC} and tabulated by \citet{ZubkoACAR}. 
From the 2018.76 $W1$ and $W2$, we derive $T_{\rmn{g}} = 649\pm8$~K, and 
$m_{\rmn{g}} = 5.2\pm0.3 \times 10^{-8} M_{\odot}$. 
With only two wavelengths, we do not have errors for the fit and those quoted 
come only from the errors on the 2018.76 and pre-outburst $W1$ and $W2$  
photometry and repeated model fits. They could be dwarfed by systematic uncertainties if the 
optical properties of the dust grains near HD~38030 differ from 
those of the laboratory sample from which they were calculated. 
The model SED is also plotted in Fig.\,\ref{FSED}. 
We see that the contribution of the dust emission to the SED in the near-IR 
is small, where the model predicts $K$ = 12.70, only 0.17 mag brighter than 
the underlying wind, and much less at $H$ and shorter wavelengths. 
This is a consequence of the relatively low grain temperature derived. 
Heating of the dust grains by the intense UV-optical radiation field 
of the WC4 and O9.7 stars prevents their survival close to the stars. 
It is only when the stellar wind material from which the grains condense 
has moved far enough away for the radiation field to be sufficiently 
diluted geometrically that the grains can survive in radiative eqilibrium, 
re-radiating in the IR. With knowledge of the stellar radiation and grain 
properties in the UV, this can be modelled: with the aid of direct 
imaging of newly formed dust around WR\,140 \citep*{MTD}, the minimum 
distance of the dust was estimated to be $\simeq$ 125~au from the stars 
\citep{W140dust}. The corresponding distance for the HD~38030 dust cloud 
also heated by WC and O stars is likely to be comparable. In any event, we 
emphasise that the dust lies a long way from the region of the WC4 star 
where the $\lambda\lambda$5801-12 \ion{C}{iv} blend is expected to form. 
For example, in the WC5 star WR\,111, the \ion{C}{iv} blend was shown to 
form within the inner 10 R* of the wind \citep[fig. 5]{Hillier89}. 
We therefore do not expect dust formation to affect the RVs measured for 
this feature in Section \ref{SRVshift} above.

Given the inference above from the light curve that dust was forming for 
not much less than a year before the 2018.76 observation, the dust mass 
derived from that observation suggests a dust-formation rate of 
$\gtrsim 5.2\pm0.3 \times 10^{-8} M_{\odot} y^{-1}$. 
If HD~38030 has a mass-loss rate typical of those derived for LMC WC4 stars 
by \citet{CrowtherWC4}, $\simeq 2 \times 10^{-5} M_{\odot} y^{-1}$, and a 
carbon abundance comparable to that of HD~32257, as indicated by the 
$\lambda$5471/$\lambda$5412 ratio, the mass of carbon flowing in the wind 
of HD~38030 would be $\simeq 10^{-5} M_{\odot} y^{-1}$. The fraction of this 
that would be available for dust formation in a wind collision depends on 
the size of the wind-collision region (WCR), which can be characterised by 
its opening angle, which is determined by the momenta of the WC and companion 
stellar winds \citep*{SBP}. 
If the opening angle of the WCR in HD~38030 was the same as that in WR\,140 
\citep[34\degr,][]{W10830} the fraction would be 0.085, providing 
$\simeq 8.5 \times 10^{-7} M_{\odot} y^{-1}$ of carbon for dust formation.
Even if the WCR in HD~38030 was an order of magnitude smaller, there would 
still be enough carbon going into the WCR to make the amount of dust
determined.

\begin{figure}                               
\centering
\includegraphics[angle=270,width=7.5cm]{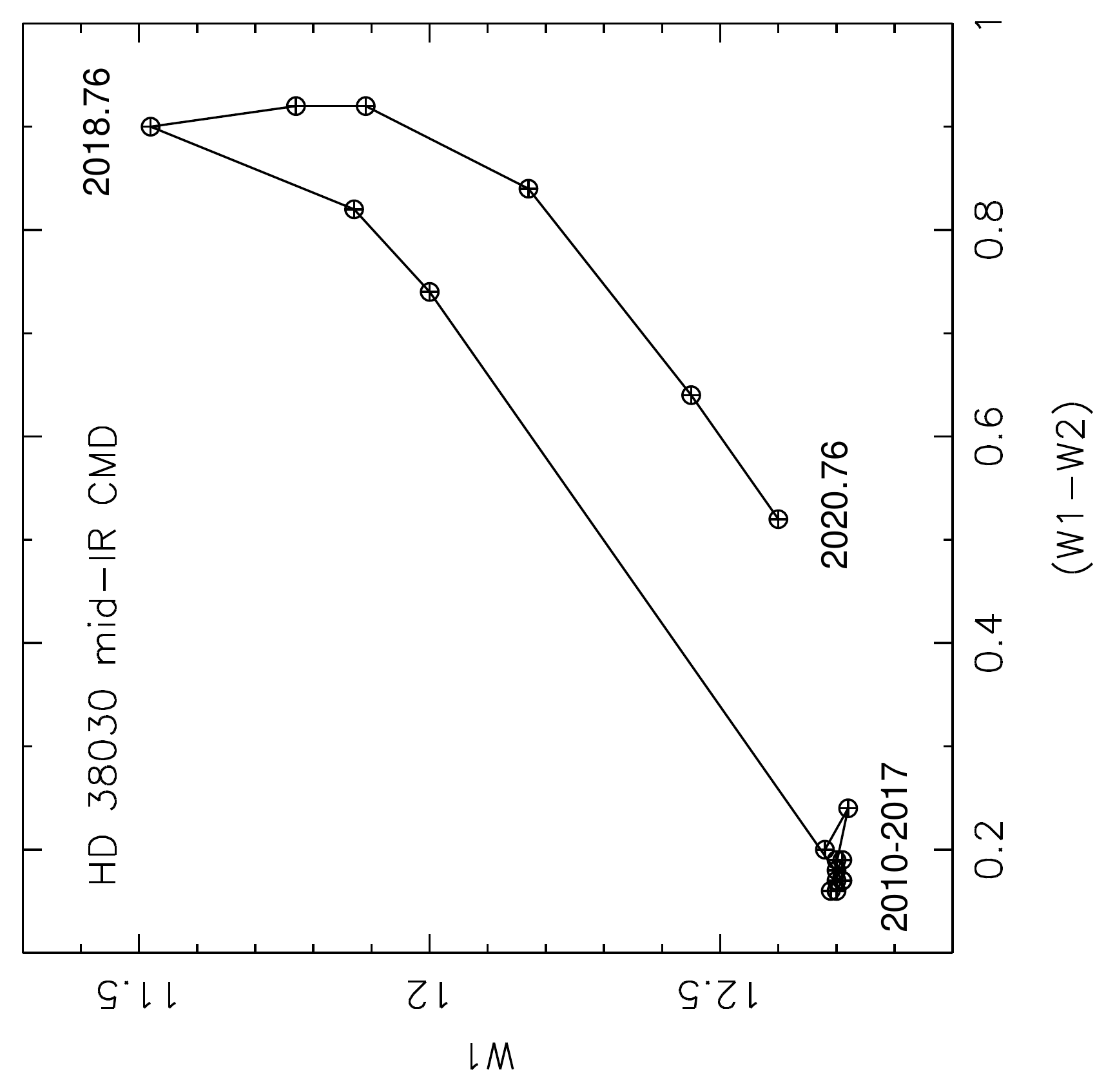}                        
\caption{Locus of HD~38030 in the mid-IR $W1$--$W2$ vs. $W1$ colour-magnitude 
diagram. The straight lines joining the observations are drawn to show the 
sequence and may skirt over more complex variation.}
\label{FCMD}
\end{figure}                                 

The behaviour during the outburst is summarised by the locus in the mid-IR 
colour-magnitude diagram (CMD) in Fig.\,\ref{FCMD}. As dust formation began, 
the star became brighter and redder in the mid-IR because the dust emission 
was redder than the stellar wind and came to dominate the emission, causing 
rapid movement of the locus to brighter $W1$ and redder $W1$--$W2$. 
This continued while new dust formed, but when formation ceased and the dust 
continued moving away from the star it cooled as the stellar radiation heating 
it was progressively diluted. This is confirmed by fitting the same model 
with the same grain emissivities, $\kappa_{\lambda}$, as above to the 2019.77 
$W1$ and $W2$ data, a year after maximum, which gives 
$T_{\rmn{g}} = 562\pm7$~K, $\sim 90$~K cooler than earlier. Again, the 
quoted error comes from the photometry and the true temperature may be 
different if the grain emissivities differ from those adopted, but the use of 
the same emissivity law for the two epochs leaves the cooling intact.
As the dust emission faded, it was redder in $W1$--$W2$ than at the 
corresponding value of $W1$ when it was rising. This contributes to the total 
emssion, resulting in the loop in the CMD. As the dust emission continues to 
fade, the locus returns slowly to the stellar wind values of $W1$ 
and $W1$--$W2$.

\subsection{The IR photometric history}

This appears to be the first recorded such event from HD 38030. Before the 
first {\em WISE} observations, the 3.2-\micron\ $N3$ magnitude observed in 
the AKARI LMC Survey was consistent with the pre-outburst $W1$ magnitudes 
(Table~\ref{TAKARI}). On the other hand, the SAGE [3.6] magnitudes observed 
in 2005 are $\sim$ 0.1 mag. fainter than the {\em WISE} $W1$. 
To see whether this indicates real variation or reflects a systematic 
difference between the datasets, we compared the SAGE [3.6] and {\em WISE} 
magnitudes for the nine LMC WC stars showing no variation in their 
{\em WISE} or NEOWISE-R data \citep{WRNEOWISE} and common to the SAGE  survey. 
We found a mean difference of [3.6]--$W1$ = 0.13$\pm$0.06. 
As this is consistent with the difference observed for HD~38030, we do not 
consider the fainter [3.6] as evidence of mid-IR variability. 
The SAGE [4.5] and {\em WISE} $W2$ are more consistent with each another. 

There are two gaps of $\sim$ 3.5~y in the mid-IR photometric history, between 
the AKARI observation in 2006.83 and the {\em WISE} All-Sky observation in 
2010.31, and between the {\em WISE} Post-Cryo observation in 2010.80 and the 
first NEOWISE-R observation in 2014.32. The question arises: could a dust 
formation episode like that observed have been completed in either of these gaps? 
Extrapolation of the fading suggests that $W1$ would not have recovered
its stellar wind level before the end of 2020 while $W2$ will not do so until about a year later, implying a duration of about 4~y. 
The $W2$ light curve therefore 
argues against a `missing' outburst in either of these gaps so that we 
conclude that the mid-IR data photometric history from 
about 2002 shows no earlier dust-formation episode. 

Taking the DENIS $K_s$ photometry at face value, there may have been an earlier 
outburst in 1996--97, indicating a period of 22~y, but the quoted errors are 
larger than those in the other data-sets. Also, there is the possibility of an 
offset between the DENIS and 2MASS data: \citet{DENISv2MASS} found a systematic 
shift of $\delta K_s$ = --0.14 (DENIS minus 2MASS) between the scales, with 
significant variation from DENIS strip to strip, which may account for the 
brighter DENIS magnitudes. 
Otherwise, on the basis of there being no evidence for dust emission 
in the first 2MASS observation, the interval between episodes, and period 
if they are periodic, must be greater than 20 yr.

\section{Discussion}

The brief dust-formation epsiode, presence of absorption lines and 
difference in RV of the 5808-\AA\ emission line from the historical work 
make HD~38030 an excellent candidate for membership of the family of 
long-period WC+O CWBs which make dust for small fractions of their orbits.
Of these, WR\,140 and WR\,19 have periods of 7.94 and 10.1~y. respectively 
but another galactic WC7+O system, WR\,125, has lately been found to be 
beginning its second observed dust-formation episode, indicating a period 
near 28.3~y \citep{WRNEOWISE}, so that the long period suggested for 
HD~38030 is not unreasonable. Further mid-IR observations on a long 
time-scale are needed to look for another episode.

The short duration of dust formation (c. 1~y) compared with the period 
($>$ 20~y) indicates that the conditions allowing this occur for a tiny 
fraction of the orbit, like WR\,140 and WR\,19, and suggests that 
HD~38030 also has a very elliptical orbit. 
Further RVs from a study such as that described by \cite{ShenarIAU} are 
needed to test this and, on a shorter time-scale, search for anti-phase 
variations in the RVs of emission and absorption lines.

The temperature derived for the dust from the 2018.76 photometry,  
$T_{\rmn{g}} = 649\pm8$~K, is lower than expected from our knowledge 
of systems like the achetypical WR\,140. 
Initially in a dust formation episode, the emission would be dominated 
by that of the hottest, newly formed grains: in the case of WR\,140,  $T_{\rmn{g}}$ = 1100~K \citep{W140dust}. 
Similarly, WR systems which make dust all the time, like the ten apparently 
constant persistent dust makers discovered in the NEOWISE-R WC stars survey 
\citep{WRNEOWISE}, have grain temperatures averaging 1112~K ($\sigma$ 198~K).
As the dust is carried away by the stellar wind, it cools because the 
stellar UV-visual radiation heating it is progressively geometrically diluted 
-- the temperature of ACAR grains in thermal equilibrium in a central 
radiation field falls off with distance $r$ from the source as 
$T_{\rmn{g}} \propto r^{-0.38}$ -- and also because the grains grow, so 
that $T_{\rmn{g}}$ had fallen to 800~K in WR\,140 a year after dust 
formation began \citep{W140dust}. 
From low-resolution IUE spectra, \citet{Niedzielski} determined a  
terminal wind velocity of 3148\kms\ for HD~38030, slightly faster than 
that \citep[2860 \kms,][]{WE2058} of WR\,140, so we expect the dispersal 
of the dust and consequent cooling in the two systems to be similar. 
The fact that the dust in HD~38030 approximately a year after condensation 
began is significantly cooler than that of WR\,140 at the same age is 
therefore probably not the result of much faster dispersal. 
This is supported by a model fit to the earlier 2018.31 observation, which 
also yielded a relatively low temperature: $T_{\rmn{g}}$ = 648$\pm$10~K. 

The dust temperatures found for HD~38030 are closer to that 
(710~K for 0.01-\micron\ ACAR grains) found from {\em Spitzer} 
IRAC [3.6] and [4.5] photometry of SPIRITS~19q, a WC4 star in NGC 2403 
which also showed a dust-formation episode peaking in 2018 \citep{Lau21}.
Like HD~38030, this system is in a metal-poor environment, and these 
differences may reflect a difference in properties of the dust or the 
circumstances of its condensation, although it is not easy to see how 
this can occur in grains formed from an element produced by the star itself. 
It would be very desirable to have IR photometry of such systems at more 
than two wavelengths to constain the dust temperature to investigate this 
further.

The collision of fast winds in a CWB may result in X-ray emission, which 
may be observable if conditions are favourable. 
In their {\em Chandra} ACS survey of WR stars in the Magellanic Clouds, 
\citet{ChandraMCs} obtained only an upper limit to the X-ray flux for 
HD~38030 but, in view of the significant variation of X-ray emission from 
the apparently similar CWBs WR\,140 \citep{Corcoran_RXTE} and WR\,19 \citep*{Naze_XWR}, further X-ray observations of HD~38030 would be worthwhile.

\section{Conclusions}

The mid-IR flux from HD~38030 rose by 1--2 mag in a year or less before 
2018.76, and then faded more slowly in 2019--20, while the $W1$--$W2$ 
colour became redder, indicative of a brief episode of circumstellar dust 
formation followed by cooling similar to those shown by the Galactic CWB 
systems WR\,140 and WR\,19. The photometric history indicates that, if 
the episodes by HD~38030 are indeed periodic, the period is greater than 
20~yr. 
 
Our new spectroscopy shows a well developed absorption-line spectrum pointing 
to the presence of a companion to the WC4 star, which we classified as O9.7 
with a tentative luminosity class of III or IV. This is supported by the 
luminosity ($M_V \simeq -5.1$) estimated from the dilution of the 
absorption lines by the WC4 continuum and photometry. 
Comparison of the $\lambda$5471 \ion{C}{iv}/$\lambda$5412 \ion{He}{ii} 
C/He diagnostic used by \citet{CrowtherWC4} with those of his sample of 
LMC WC4 stars shows that the C/He abundance ratio in HD~38030 is at the 
high end of those in the sample, suggesting by comparison with the results 
of his modelling, C/He $\simeq$ 0.35 in HD~38030.
Comparison of the radial velocity of the $\lambda$5801--12 \ion{C}{iv} 
blend in our 2019 and 2020 spectra measured in the same way as that of 
the same feature by \citet{Bartzakos} in 1984 and 1993 show a significant 
(77$\pm$12~\kms) shift. 
Assuming that this shift is not caused by variations in the shape of 
the \ion{C}{iv} emission line profile, it strengthens the view that the 
WC4 and O9.7 stars are members of a binary system, a colliding 
wind binary.

Modelling the dust emission yielded a mass, 
$m_{\rmn{g}} = 5.4\pm0.3 \times 10^{-8} M_{\odot}$, indicating a formation 
rate consistent with the amount of carbon expected to be flowing into the WCR 
and available for condensation. The grain temperatures at maximum and when
the IR flux was rising were, however, lower than those observed from the 
archetypical system WR\,140 at similar stages in the evolution of its dust cloud.
The temperature was closer to that (710~K) found by \citet{Lau21} from 
{\em Spitzer} IRAC [3.6] and [4.5] photometry for SPIRITS~19q, a WC4 star 
which showed a dust-formation episode peaking in 2018 and, like HD~38030, 
also lies in a metal-deficient environment (NGC 2403).

\section*{Acknowledgments}
This paper includes data gathered at the 6.5-meter Magellan telescopes at Las 
Campanas Observatory, Chile. 
The paper makes use of data products from the Near-Earth Object Wide-field 
Infrared Survey Explorer (NEOWISE-R), which is a project of the Jet Propulsion 
Laboratory/California Institute of Technology. NEOWISE-R is funded by the National 
Aeronautics and Space Administration. Data were retrieved from the NASA/ IPAC 
Infrared Science Archive, which is operated by the Jet Propulsion Laboratory, 
California Institute of Technology, the VISTA Science Archive, which is operated 
by the Wide Field Astronomy Unit of the Institute for Astronomy, University of 
Edinburgh, and the VizieR catalogue access tool, operated by the CDS, Strasbourg.
PMW is grateful to the Institure for Astronomy for continued hospitality 
and access to the facilities of the Royal Observatory Edinburgh. 
We would like to thank the referee for a careful report.

\section*{Availability of data}
The infrared photometry is publicly available from the references cited. The 
spectra will be shared on reasonable request to the corresponding author.

\bibliographystyle{mnras}
\bibliography{HD38030}           

\bsp	
\label{lastpage}

\end{document}